\documentclass[aps,prl,twocolumn,superscriptaddress]{revtex4}
\usepackage{graphicx}
\usepackage{epstopdf}
\usepackage{latexsym}
\usepackage{amssymb}
\usepackage{amsmath}
\usepackage{amsfonts}
\usepackage{subfigure}
\usepackage{bm}
\usepackage{multirow}

\newcommand{\vect}[1]{{\bm{#1}}}
\newcommand{\eqnref}[1]{Eq.\,\eqref{#1}}
\newcommand{\figref}[1]{Fig.\,\ref{#1}}

\begin{document}

\title{Superfluidity of Bosons in Kagome Lattices with Frustration}
\author{Yi-Zhuang You}
\affiliation{Institute for Advanced Study, Tsinghua University, Beijing, 100084, China}
\author{Zhu Chen}
\affiliation{Institute for Advanced Study, Tsinghua University, Beijing, 100084, China}
\author{Xiao-Qi Sun}
\affiliation{Institute for Advanced Study, Tsinghua University, Beijing, 100084, China}
\author{Hui Zhai}
\email{hzhai@tsinghua.edu.cn}
\affiliation{Institute for Advanced Study, Tsinghua University, Beijing, 100084, China}
\date{\today }

\begin{abstract}

In this letter we consider spinless bosons in a Kagome lattice with nearest-neighbor hopping and on-site interaction, and the sign of hopping is inverted by insetting a $\pi$ flux in each triangle of Kagome lattice so that the lowest single particle band is perfectly flat. We show that in the high density limit, despite of the infinite degeneracy of the single particle ground states, interaction will select out the Bloch state at the $K$ point of Brillouin zone for boson condensation at the lowest temperature. As temperature increases, the single boson superfluid order can be easily destroyed, while an exotic triple-boson paired superfluid order will remain. We establish that this trion superfluid exists in a broad temperature regime until the temperature is increased to the same order of hopping and then the system turns into normal phases. Finally we show that time of flight measurement of momentum distribution and its noise correlation can be used to distinguish these three phases. 

\end{abstract}

\maketitle

\begin{figure}[btp]
\begin{center}
\includegraphics[width=0.28\textheight]{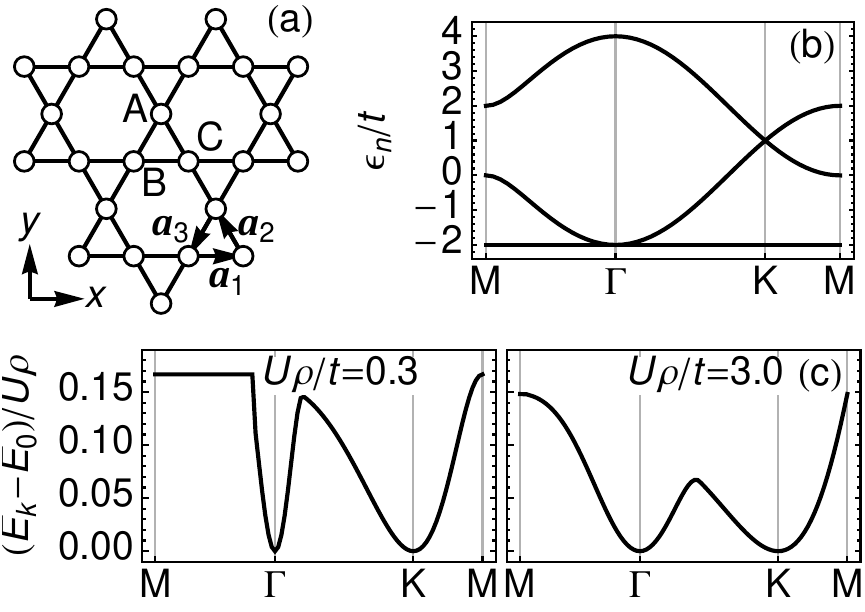}
\caption{(a) Kagome lattice, partitioned into A, B, C sublattices. (b) Band structure with inverted hopping sign, where $\Gamma=(0,0)$, $K=(2\pi/3,0)$ and $M=(\pi/2,\pi/2\sqrt{3})$. (c) Mean field energy landscape for different $U\rho/t$. Energy shifted by $E_0=-2t+U\rho/3$.\label{Kagome}}
\end{center}
\end{figure}

Flat band models have attracted considerable theoretical interests recently \cite{Altman,photonic,ETang,SunGu,NSCM,WangRan,Ashvin1,Ashvin2,Wu,YFWang1,YFWang2,Cooper,Kou,CWu}, because the single particle energies are degenerate inside the flat band, therefore interactions play a dominant role in the many-body system and lead to many interesting quantum phases \cite{Ashvin1,Ashvin2,Wu,YFWang1,YFWang2,Cooper,Kou}. Among many different physical realizations of lattice models with flat band, Kagome lattices with only nearest neighbor hopping is perhaps one of the simplest. Recently, such a model has been realized experimentally using optical lattices by the Berkeley group \cite{Stamper-Kurn}. However, the flat band in a normal Kagome lattice is the highest band. By fast shaking the optical lattices, one can invert the sign of hopping, which has also been demonstrated experimentally for triangular optical lattices \cite{sengstock}. This technique can be applied straightforwardly to Kagome lattices. When the sign of hopping is inverted, it is equivalent to inserting a $\pi$ flux in each triangle of the Kagome lattice \cite{note}, and the flat band becomes the lowest band, as shown in Fig.\,\ref{Kagome}.

In this letter we consider spinless bosons with on-site interaction on the Kagome lattice with sign-inverted hopping, such that the boson condensation is frustrated\cite{CWu}. Here we focus on the high density superfluid regime, since those strongly correlated phases like Mott insulator\cite{Ashvin1,Ashvin2}, Wigner crystal\cite{Wu} and quantum Hall state\cite{YFWang1,YFWang2} usually occurs in low-density regime. This high density limit corresponds to a real system with a Kagome optical lattice in the $xy$ plane and weak confinement potential along the $\hat{z}$ direction. Therefore, each site is in fact a tube inside which bosons form a quasi-condensate, and tunneling couples different tubes into a two-dimensional Josephson array described by a boson Hubbard model. 

Here we shall discuss whether and how bosons can condense in the flatband, and what exotic type of superfluid occurs at finite temperature. The main findings are

(\textbf{1}) As the temperature increases, the system exhibits three different phases: a $K$-point ($\sqrt{3}\times\sqrt{3}$) superfluid phase, with bosons condensed to the single-particle Bloch state at the momentum $K$ point; an exotic ``trion superfluid'' phase, with triple-boson (quasi-)long-range order in the absence of single-boson ordering; and a normal phase of thermal boson gas without any order.

(\textbf{2}) The three phases are separated by two phase transitions at very different temperature scales. The higher one is of the hopping energy scale, while the lower one is three orders of magnitude smaller. Thus we predict a large temperature window in which the ``trion superfluid'' phase can be observed in experiments.

(\textbf{3}) We show that the time of flight (TOF) detection of momentum distribution can distinguish these three phases, and the third-order noise correlation provides a more direct evidence of the ``trion superfluid" phase.

It has been a long term effort to search for exotic bosonic state without single-boson superfluid order. In previous studies, two-boson paired superfluids (or charge-$4e$ superconductor as paired Cooper pairs) have been proposed in various systems \cite{Kivelson,FFLO1,FFLO2,FFLO3,FZhou,XuMoore,Ashvin4,Lamacraft,SOC}. 
As far as we know, it is the first time that a triple-boson paired superfluid is proposed, and the underlying physics, i.e. the frustrated hopping in Kagome geometry, is also different from that of previous examples. Moreover, unlike previous proposals where unconventional superfluids always take place at very low temperature, here the trion superfluid exists in relatively high temperature regime, and thus, it is easier for experimental realization. These results may also be generalized to interacting bosons in other flat band models with geometric frustration.

\emph{Band Structure and Mean-Field.} The boson Hubbard model on the Kagome lattice with $\pi$ flux in each triangle is given by $\hat{H} = \hat{H}_t + \hat{H}_U$, with the hopping term $\hat{H}_t = t\sum_{\langle i j\rangle}\hat{b}_i^\dagger \hat{b}_j+h.c.$, and the interaction term  $\hat{H}_U = U\sum_i \hat{n}_i(\hat{n}_i-1)-\mu\sum_i \hat{n}_i$, where $\hat{n}_i=\hat{b}_i^\dagger \hat{b}_i$. Here $t$ is positive \cite{note}.
In the momentum space, the hopping Hamiltonian reads $\hat{H}_t=\sum _\vect{k} \hat{b}_\vect{k}^{\dagger }h(\vect{k})\hat{b}_\vect{k}$ with $\hat{b}_\vect{k}=(\hat{b}_{\vect{k}\text{A}},\hat{b}_{\vect{k}\text{B}},\hat{b}_{\vect{k}\text{C}})^\intercal$ and
\begin{equation}
h(\vect{k})=2t\left(
\begin{array}{ccc}
 0 & \cos  k_3 & \cos  k_2 \\
 \cos  k_3 & 0 & \cos  k_1 \\
 \cos  k_2 & \cos  k_1 & 0
\end{array}
\right),
\end{equation}
where $k_i\equiv \vect{k}\cdot\vect{a}_i$, and $\vect{a}_1=(1,0)$, $\vect{a}_2=(-\frac{1}{2},\frac{\sqrt{3}}{2})$, $\vect{a}_3=(-\frac{1}{2},-\frac{\sqrt{3}}{2})$ as shown in \figref{Kagome}(a).  Its lowest band is perfectly flat with a quadratic band touching at $\Gamma$ point, as shown in \figref{Kagome}(b).

Because of the infinite degeneracy of the single particle ground state, free bosons can not condense. 
At the mean-field level, we first consider the single particle state with translational symmetry labelled by momentum $\vect{k}$. Taking the mean-field ansatz $\langle \hat{b}_{\vect{k}}\rangle = z\equiv(z_{\text{A}},z_{\text{B}},z_{\text{C}})^\intercal$ with the normalization condition $z^\dagger z=\rho$, where $\rho$ is the boson filling per unit cell, we can minimize the mean-field energy function $E_{\vect{k}}[z]=z^\dagger h(\vect{k})z+U(|z_{\text{A}}|^4+|z_{\text{B}}|^4+|z_{\text{C}}|^4)$ with respect to $z$, and denote the optimal energy as $E_\vect{k}=\min_{z}E_\vect{k}[z]$. This mean-field energy landscape in \figref{Kagome}(c) shows that the single particle degeneracy is lifted by the Hartree energy. $\Gamma$ and $K$ points are selected out to be the energy minimum. This is because inside the flat band, the single particle wave functions have equal amplitudes among the three sublattices only at $\Gamma$ and $K$ points, and their uniform densities are favored by the repulsive interaction.

\begin{figure}[tbp]
\begin{center}
\includegraphics[width=0.23\textheight]{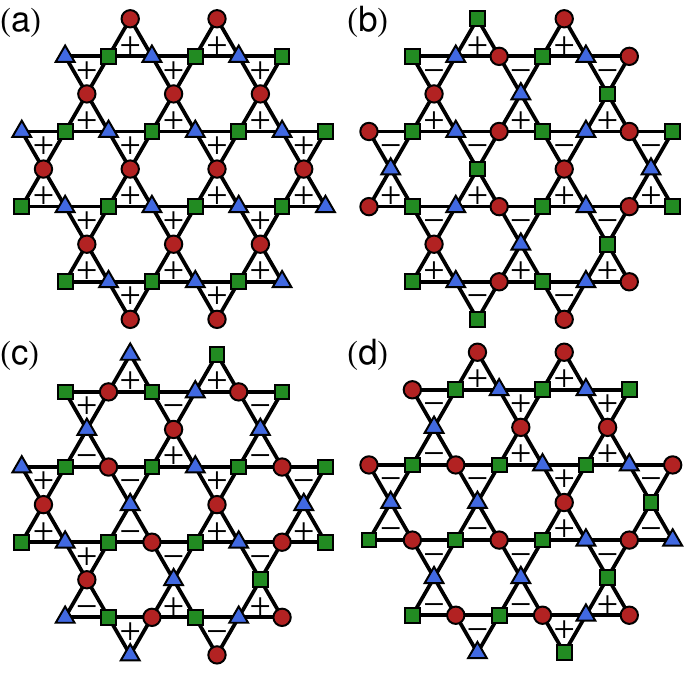}
\caption{(Color online.) (a-b) Phase configurations of the condensates at  $\Gamma$ point (a) and $K$ point (b), (c-d) random ``3-color arrangement''. Phase $\theta_i$ is denoted by colors: red circle  $=0$, green square $=2\pi i/3$, blue triangle $=-2\pi i/3$. ``$\pm$'' mark out the vorticity around each triangle. (a) $\psi_{\Gamma}$ is a ``vorticity ferromagnetic" state and (b) $\psi_{K}$ is a ``vorticity antiferromagnetic" state.}\label{config}
\label{fig:configuration}
\end{center}
\end{figure}

The Bloch wave functions for $\Gamma$ and $K$ points is determined by minimizing mean-field energy,
\begin{equation}
\begin{split}
\psi_{\Gamma}(\vect{r}_i)&=\tfrac{1}{\sqrt{3}}(1,e^{\pm 2\pi i/3},e^{\mp 2\pi i/3})^\intercal,\\
\psi_{K}(\vect{r}_i)&=\tfrac{1}{\sqrt{3}}e^{i\vect{k}_K\cdot \vect{r}_i}(1,-1,-1)^\intercal,
\end{split}
\end{equation}
with $\vect{k}_K=(2\pi/3,0)$. In the real space, both two wave functions satisfy two conditions: (i) their densities are uniform, which minimize the mean-field interaction energy and (ii) their phases follow the ``3-color arrangement", meaning that each pair of two neighboring sites take different phases out of $1$, $e^{2\pi i/3}$ and $e^{-2\pi i/3}$, as depicted in \figref{config}, such that the kinetic energy is also minimized. The mean-field energy is minimized to $E_0=-2t+U\rho/3$, as long as both conditions (i) and (ii) are satisfied, while the translation invariance is not a necessary condition and can be released. Then one can find extensive numbers of state without translation invariance but satisfying both (i) and (ii), as exemplified in \figref{config}(c-d).

\emph{Quantum Fluctuation and Order from Disorder.} These extensive number of degenerate mean-field states can be further lifted by quantum fluctuations, because the zero-point energy (ZPE) of Bogoliubov phonons is different for different mean-field state. For a given mean-field configuration $\langle \hat{b}_i\rangle=\sqrt{\rho}e^{i\theta_i}$, its Bogoliubov Hamiltonian is 
\begin{equation}
\begin{split}
H[\theta_i]=&t\sum _{\langle i j\rangle }\left( \hat{b}_i^{\dagger }\hat{b}_j+h.c.\right)-\mu\sum_i \hat{b}_i^\dagger \hat{b}_i\\
&+U\rho \sum _i \left(2\hat{b}_i^\dagger \hat{b}_i+e^{2i\theta_i} \hat{b}_i^{\dagger }\hat{b}_i^{\dagger }+h.c.\right),
\end{split}
\end{equation}
where $\mu=-2t+2U\rho/3$. Diagonalization of $H[\theta_i]$ leads to $H[\theta_i]=\sum_m \omega_m[\theta_i] \left(\hat{\gamma}_m^\dagger\hat{\gamma}_m + 1/2\right)$, with $\hat{\gamma}_m$ being Bogoliubov boson operator. Thus the ZPE associated with this given $\{\theta_i\}$ configuration equals to $\Lambda[\theta_i]=\frac{1}{2}\sum_m \omega_m[\theta_i]$ by setting $\hat{\gamma}_m^\dagger\hat{\gamma}_m=0$.

\begin{figure}[tbp]
\begin{center}
\includegraphics[width=0.30\textheight]{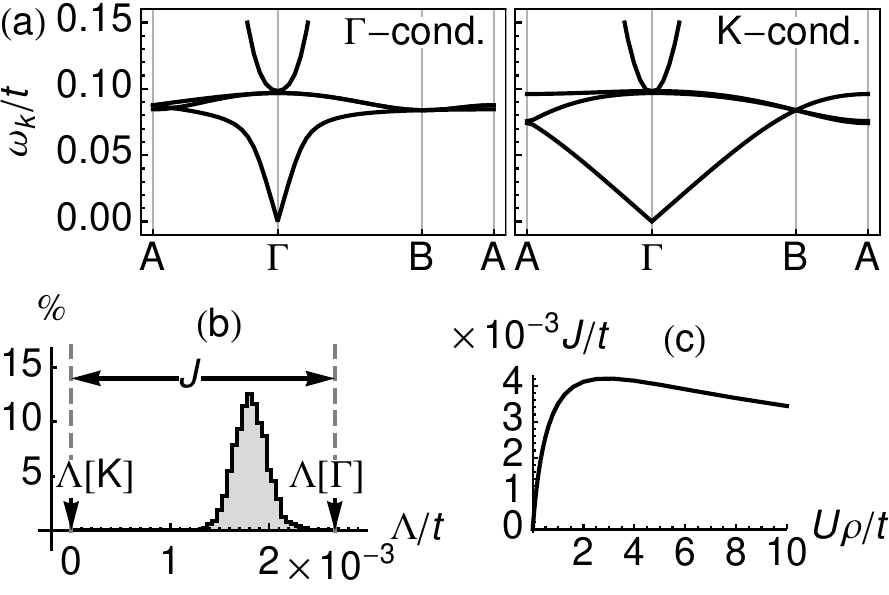}
\caption{(a) Bogoliubov spectra for $\Gamma$ point (left) and $K$ point (right) condensates at $U\rho/t=0.3$. Momentum points are defined as $A=(\pi/3,0)$ and $B=(\pi/3,\pi/3\sqrt{3})$. (b) Distribution of the zero-point energy (ZPE) $\Lambda$ of the degenerate mean-field configurations. Two dashed lines denote the ZPE for $K$ and $\Gamma$ point condensates. The vertical axes is the probability for the particular ZPE. (c) $J/t$ v.s. $U\rho/t$ plot. \label{Bogoliubov}}
\end{center}
\end{figure}

For condensates at $\Gamma$ and $K$ points, the Bogoliubov excitations have quantum number $\vect{k}$ and they have well defined dispersion as shown in \figref{Bogoliubov}(a). One finds that the sound velocity $c$ of $K$ point condensate is smaller that that of $\Gamma$ point condensate, hence $K$ point condensate has lower ZPE, as $\Lambda\sim c^2$. This is also evidenced from the mean field energy landscape as shown in \figref{Kagome}(c), where energy landscape changes less rapidly nearby $K$ point than that nearby $\Gamma$ point, indicating softer Goldstone mode and lower ZPE.

For a generic mean-field state in the degenerate manifold, the Bogoliubov spectrum has no well defined dispersion because of the absence of translation symmetry. Here we randomly sample $4000$ configurations on a 240-site Kagome lattice with uniform density and satisfying the ``3-color arrangement", and then calculate their ZPE's numerically. We find that their ZPE's all rest between $\Gamma$ point condensate and $K$ point condensate, i.e. $\forall \{\theta_i\}:\Lambda[K]\leq\Lambda[\theta_i]\leq\Lambda[\Gamma]$, as seen from \figref{Bogoliubov}(b). So the degeneracy can be completely removed via the order-by-disorder mechanism. At sufficiently low temperature, bosons will condense to $K$ point (or its symmetry related points).

\emph{Thermal Fluctuation and Phase Diagram.} 
Because all the ZPE's are in the range from $\Lambda[K]$ to $\Lambda[\Gamma]$, it is natural to introduce the energy scale for zero-point fluctuation as $J=\Lambda[\Gamma]-\Lambda[K]$. Beyond this energy scale, the condensation at $K$ point will be destroyed.
As shown in \figref{Bogoliubov}(c), $J$ will vanish in both small and large $U$ limit. The asymptotic behavior goes like $J\propto U\rho$ for $U\rho\ll t$ and $J\propto t^{3/2}(U\rho)^{-1/2}$ for $U\rho\gg t$. The maximum of $J$ is achieved around $U\rho/t=3$. It is remarkable to find that even the maximum value of $J$ is three orders of magnitude smaller than $t$. It is still quite challenging to reach such low temperature in current cold atom experiments.

However, the conventional Bose condensate at the lowest temperature is not the most interesting phase. The most interesting phase in this system exists at the temperature regime $k_{\text{B}}T>J$. In this regime, the system will enter a thermal mixed state in which all mean-field configurations satisfying condition (i) and (ii) mentioned above are almost equally populated. In this thermal mixed state, for each site the condensate phase can take $\theta_i=0,\pm2\pi/3$ with equal probabilities. Thus, the single boson correlation will become short-ranged, i.e. $\langle \hat{b}^\dag_i \hat{b}_j\rangle = \rho_s \langle e^{-i\theta_i}e^{i\theta_j}\rangle \to 0$. However,  as long as the ``3-color arrangement" is satisfied, at every site  $e^{3i\theta_i}\equiv1$ is always hold. Thus, the triple bosons operator can still possess long-range correlation, i.e. $\langle \hat{b}_i^{\dag 3} \hat{b}_j^3\rangle = \rho_s^3 \langle e^{-3i\theta_i} e^{3i\theta_j}\rangle\to\rho_s^3$ \cite{Huse}. This trion superfluid (trion SF) supports many interesting properties, such as $1/3$ fractionalized vortices, which will be studied in the future. 

Since the ``3-color arrangement" is enforced by the kinetic energy, thus, bosons will condense in triples as long as $T<t$. When $T$ is increased to $\sim t$, the long wave-length fluctuation of the $U(1)$ phase will lead to a Kosterlitz-Thouless phase transition from the trion SF to the normal state. This transition temperature can be estimated from the superfluid stiffness, which can be calculated from the free energy response to the phase twist. Let $b_i=w_i e^{i\theta_i}e^{i \vect{q}\cdot\vect{r}_{i}}$, where $e^{i\vect{q}\cdot\vect{r}_i}$ is the applied phase twist. In the temperature regime $T\gg J$, we can ignore the zero-point energy, and the energy functional for a given $\{\theta_i\}$ configuration reads $E_\vect{q}[\theta_i,w_i]=t\sum_{\langle i j\rangle} (w_i^*w_j e^{-i(\theta_i-\theta_j)} e^{-i\vect{q}\cdot(\vect{r}_i-\vect{r}_j)}+h.c.)+U\sum_i |w_i|^4$, and because all the configurations must be taken into account under thermal average, the averaged energy functional reads 
\begin{align}
&\mathcal{E}_\vect{q}[w_i]=\sum_{\{\theta_i\}}E_\vect{q}[\theta_i,w_i] \nonumber\\
&=-\frac{t}{2}\sum_{\langle i j\rangle} (w_i^*w_j e^{-i\vect{q}\cdot(\vect{r}_i-\vect{r}_j)}+h.c.) +U\sum_i |w_i|^4 \label{trion_eng}
\end{align}
It is interesting to note that \eqnref{trion_eng} is the same of a mean-field energy of bosons in the Kagome lattice {\it without} frustration, because the sign of hopping is now inverted back. Mathematically, it is because taking thermal averages lead to $\langle e^{i(\theta_i-\theta_j)}\rangle=-1/2$. Since the kinetic energy frustration is released, trion SF will have a finite stiffness. It can be obtained by minimizing $\mathcal{E}_\vect{q}[w_i]$ with respect to $w_i$ and expand the optimal energy in terms of small $\vect{q}$, which leads to $\min_{\{w_i\}}\mathcal{E}_\vect{q}[w_i]\simeq -2t+\frac{U\rho}{3}+\frac{1}{2}t\vect{q}^2$. Therefore, the stiffness is $t$, which gives an estimate of the Kosterlitz-Thouless transition temperature as $T_\text{KT}=t/\pi$. For Rb atoms, $t$ is around $15$nK with a moderate lattice depth, and thus $T_{\text{KT}}$ is a few nK. 

\begin{figure}[tbp]
\begin{center}
\includegraphics[width=0.32\textheight]{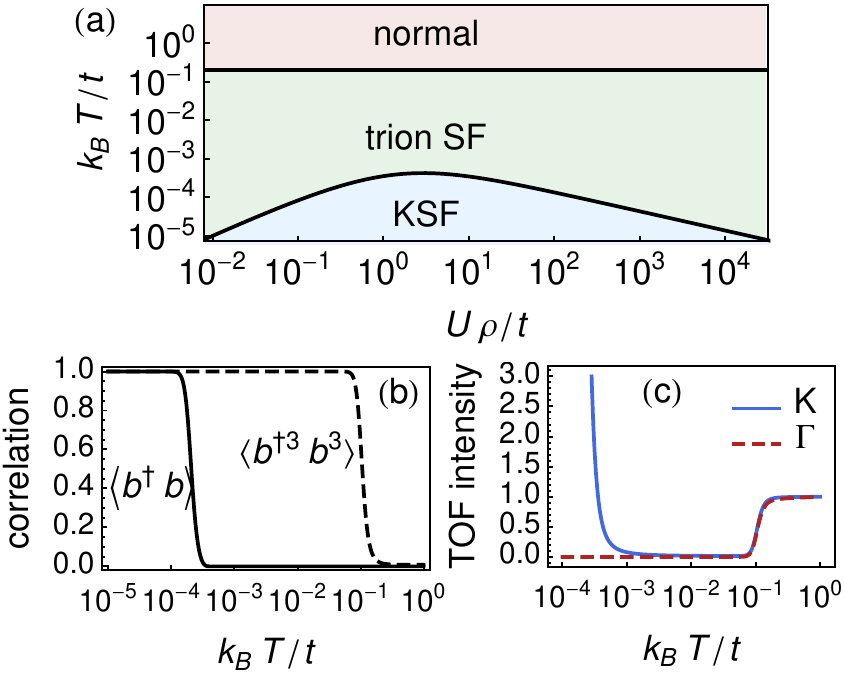}
\caption{(Color on line.) (a) Phase diagram. It contains three phases: the conventional superfluid phase with boson condensation at $K$ point (KSF), trion superfluid (trion SF) phase, and the normal phase. 
(b) The long-range correlation for $\langle \hat{b}_i^\dag \hat{b}_j\rangle$ and $\langle \hat{b}_i^{\dag 3} \hat{b}_j^3\rangle$ as a function of temperature $k_{\text{B}}T/t$. 
(c) TOF intensity at $\Gamma$ and $K$ point as a function of temperature $k_{\text{B}}T/t$. (b) and (c) are calculated at $U\rho/t=1$. \label{phase}}
\end{center}
\end{figure}

With the analysis above, we reach the phase diagram as shown in \figref{phase}(a). In \figref{phase}(b), we show the one-boson and three-boson correlation function as a function of temperature for a fixed $U\rho/t$. The one-particle correlation $\langle \hat{b}^\dag_i \hat{b}_j\rangle$ is calculated as $\rho_s\sum_{\{\theta_i\}}e^{-i(\theta_i-\theta_j)}e^{-\Lambda[\theta_i]/T}$, where the configuration summation can be restricted in the ``3-color arrangement'' patterns, since other configurations cost energy of the order $t$ and their contribution is negligible at low temperature around the order $J$. While on the other hand, when we calculate the three-boson correlation $\langle \hat{b}_i^{\dag 3} \hat{b}_j^3\rangle=\rho_s^3 \sum_{\{\theta_i\}} e^{-3i(\theta_i-\theta_j)}e^{-E[\theta_i]/T}$, the summation goes over all configurations $\{\theta_i\}$, and the energy is given by $E[\theta_i]=2t \sum_{\langle ij \rangle}\cos{(\theta_i-\theta_j)}$, where we have ignored the ZPE as it is negligible compared to $E[\theta_i]$. \figref{phase}(b) shows that there indeed exists a large temperature window there the one-boson correlation function vanishes while the three-boson correlation function remains finite. 

\begin{figure}[tbp]
\begin{center}
\includegraphics[width=0.26\textheight]{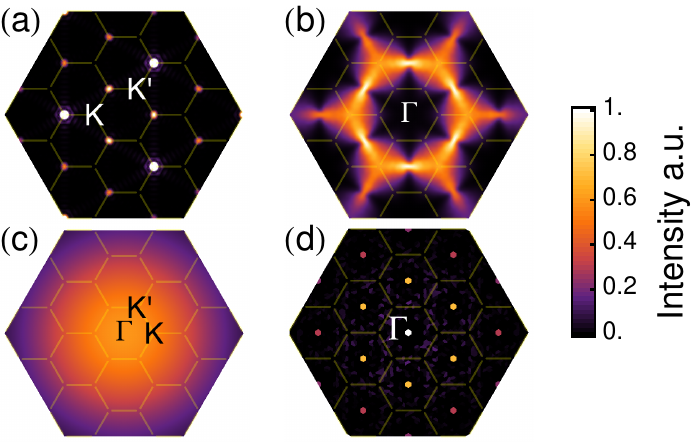}
\caption{(Color on line.)  (a-c) TOF image of momentum distribution for three phases: (a) KSF, (b) trion SF, (c) normal. The thin lines of honeycomb trace out the Brillouin zone boundary. For (b) and (c), the image is averaged over many times of TOF experiments. (d) The profile of $C(\vect{k})$ simulated in the trion SF phase by analyzing the noise correlation among 100 TOF images.}\label{TOF}
\end{center}
\end{figure}

\emph{Detection.} Finally we show that the difference between these three phases can be detected by a straightforward measurement of momentum distribution via TOF image \cite{note2}. When bosons condense into $K$ points of the Brillouin zone, TOF image will display sharp Bragg peaks at $K$ points (or its equivalent $K^\prime$ points) in the reciprocal lattice, as shown in \figref{TOF}(a). Moreover, due to the interference effect from the phase structure satisfying the ``3-color-arrangement", it is easy to show that the strongest peaks do not appear in the first Brillouin zone, but instead at its reciprocal lattice vector points in the second Brillouin zone. As temperature increases to the trion SF phase, all the 3-color arrangements are thermally mixed, and the Bragg peak at $K$ point disappears as the single-boson superfluid order vanishes. However, in contrast to normal state, the TOF image displays nontrivial features as shown in \figref{TOF}(b). A large honeycomb structure appears and the hexagon is $4$ times the area of the first Brillouin zone, and the intensity at $\Gamma$ point is always zero, because for zero-momentum component of the Fourier transformation is exactly cancelled out due to the ``3-color arrangement". At the highest temperature normal state, the TOF image becomes a featureless Gaussian and the intensity at $\Gamma$ point becomes the maximum, as shown in \figref{TOF}(c). Thus, we predict that the intensity at $K$ point rapidly decreases at the transition from $K$-point condensate to trion SF, and the intensity of $\Gamma$ point rapidly increases at the transition from trion SF to normal state, as shown in \figref{phase}(c). 

Besides, the trion SF order can also be probed by analyzing the three-point noise correlation of the TOF images. This requires repeating the TOF experiment many times to obtain the noise signal $\delta n_\vect{k}=n_\vect{k}-\langle n_\vect{k}\rangle$  for each image like Fig. \ref{TOF}(b) \cite{Demler,Bloch}. We propose that the triple-boson long-range correlation can be observed from the three-point noise correlation $C(\vect{k})=\sum_{\vect{k}_1+\vect{k}_2+\vect{k}_3=\vect{k}}\langle \delta n_{\vect{k}_1}\delta n_{\vect{k}_2}\delta n_{\vect{k}_3}\rangle$, where it is important to collect information from all the points satisfying $\vect{k}_1+\vect{k}_2+\vect{k}_3=\vect{k}$ into $C(\vect{k})$. It is straightforward to show that this noise correlation displays sharp feature only in the trion SF phase, and the triple-boson correlation is directly related to the noise correlation via $\langle \hat{b}_i^{\dagger3} \hat{b}_j^3\rangle\propto\sum_{\vect{k}_{1,2,3}}\langle \delta n_{\vect{k}_1}\delta n_{\vect{k}_2}\delta n_{\vect{k}_3}\rangle e^{-i(\vect{k}_1+\vect{k}_2+\vect{k}_3)(\vect{r}_i-\vect{r}_j)}$. Therefore, $C(\vect{k})$ shows sharp peaks at $\Gamma$ points in the trion SF phase, as in \figref{TOF}(d), providing direct evidence for  the long-range correlation of $\langle \hat{b}_i^{\dagger3} \hat{b}_j^3\rangle$.

{\it Acknowledgment}: We would like to thank T.-L. Ho and H. Yao for helpful discussions. This work is supported by Tsinghua University Initiative Scientific Research Program, NSFC Grant No. 11004118 and No. 11174176, and NKBRSFC under Grant No. 2011CB921500.


\begin{thebibliography}{99}

\bibitem{Altman}
S. D. Huber and E. Altman, Phys. Rev. B {\bf 82}, 184502 (2010) 	
\bibitem{photonic}
S. Endo, T. Oka, and H. Aoki, Phys. Rev. B {\bf 81}, 113104 (2010) 
\bibitem{ETang}
E. Tang, J.-W. Mei, and X.-G. Wen, Phys. Rev. Lett. {\bf 106}, 236802 (2011) 
\bibitem{SunGu}
K. Sun, Z.-C. Gu, H. Katsura, S. Das Sarma, Phys. Rev. Lett. {\bf 106}, 236803 (2011) 
\bibitem{NSCM}
T. Neupert, L. Santos, C. Chamon, C. Mudry, Phys. Rev. Lett. {\bf 106}, 236804 (2011) 
\bibitem{WangRan}
F. Wang, Y. Ran, Phys. Rev. B {\bf 84}, 241103(R) (2011) 
\bibitem{Ashvin1}
S. A. Parameswaran, I. Kimchi, A. M. Turner, D. M. Stamper-Kurn, A. Vishwanath, arXiv:1206.1072v1.
\bibitem{Ashvin2}
I. Kimchi, S. A. Parameswaran, A. M. Turner, A. Vishwanath, arXiv:1207.0498v1.
\bibitem{Wu}
C. Wu, D. Bergman, L. Balents, and S. Das Sarma, Phys. Rev. Lett. {\bf 99}, 070401 (2007) 
\bibitem{YFWang1}
Y.-F. Wang, H. Yao, Z.-C. Gu, C.-D. Gong, D. N. Sheng, Phys. Rev. Lett. {\bf 108}, 126805 (2012) 
\bibitem{YFWang2}
Y.-F. Wang, H. Yao, C.-D. Gong, D. N. Sheng, arXiv:1204.1697v1.
\bibitem{Cooper}
G. M\"oller and N. R. Cooper, Phys. Rev. Lett. {\bf 108}, 045306 (2012) 
\bibitem{Kou}
X.-H. Zhang, S.-P. Kou, arXiv:1205.6641
\bibitem{CWu}
Z. Cai, Y. Wang, C. Wu, Phys. Rev. B {\bf 86}, 060517(R) (2012);  G.-W. Chern, C. Wu, arXiv:1204.6019.
\bibitem{Stamper-Kurn}
G.-B. Jo, J. Guzman, C. K. Thomas, P. Hosur, A. Vishwanath, and D. M. Stamper-Kurn, Phys. Rev. Lett. {\bf 108}, 045305 (2012) 
\bibitem{sengstock}
J. Struck, C. \"Olschl\"ager, R. Le Targat, P. Soltan-Panahi, A. Eckardt, M. Lewenstein, P. Windpassinger, and K. Sengstock, Science {\bf 333}, 996 (2011) 
\bibitem{note}
This is gauge equivalent to the situation that one of the three hopping signs (from A site to B site, from B site to C site or from C site to A site) is inverted, and such a situation may be easier for experimental implementation. 
\bibitem{Kivelson}
E. Berg, E. Fradkin and S. A. Kivelson, Nature Phys. {\bf 5}, 830 - 833 (2009) 
\bibitem{FFLO1}
L. Radzihovsky and A. Vishwanath, Phys. Rev. Lett. {\bf 103}, 010404 (2009) 
\bibitem{FFLO2}
L. Radzihovsky, Phys. Rev. A {\bf 84}, 023611 (2011) 
\bibitem{FFLO3}
D. F. Agterberg and H. Tsunetsugu, Nature Phys. {\bf 4}, 639 (2008) 
\bibitem{FZhou}
F. Zhou, Phys. Rev. Lett. {\bf 87}, 80401 (2001)
\bibitem{XuMoore}
S. Mukerjee, C. Xu, and J. E. Moore, Phys. Rev. Lett. {\bf 97}, 120406 (2006) 
\bibitem{Ashvin4}
D. Podolsky, S. Chandrasekharan, and A. Vishwanath, Phys. Rev. B {\bf 80}, 214513 (2009) 
\bibitem{Lamacraft}
A. J. A. James and A. Lamacraft, Phys. Rev. Lett. {\bf 106}, 140402 (2011) 	
\bibitem{SOC}
C.-M. Jian and H. Zhai, Phys. Rev. B {\bf 84}, 060508(R) (2011) 
\bibitem{Huse}
Similar argument has been presented in spin model, see D. A. Huse and A. D. Rutenberg, Phys. Rev. B {\bf 45}, 7536–7539 (1992).

\bibitem{note2}
For the case of flipping the hopping sign along only one type of bounds, its TOF image will be shifted by one reciprocal lattice vector.
\bibitem{Demler}
E. Altman, E. Demler, M. D. Lukin, Phys. Rev. A {\bf 70}, 013603 (2004).
\bibitem{Bloch}
S. Fölling, F. Gerbier, A. Widera, O. Mandel, T. Gericke, I. Bloch, Nature {\bf 434}, 481-484 (2005).

\end{thebibliography}
\end{document}